\documentclass[12pt,preprint]{aastex}
\begin{document}
\title{Deep meridional circulation below the solar convective envelope}

\author{K. M. Hiremath}

\affil{Indian Institute of Astrophysics, Bangalore-560034, India}

\email{hiremath@iiap.res.in}

\begin{abstract}
With reasonable assumptions and approximations, we compute the velocity
of the meridional flow $U$ in the convective envelope by modified
Chandrasekhar's (1956) MHD equations.
The analytical solution of such a modified equation is found to be
$U(x,\mu) = \sum_{n=0}^\infty \bigl[u1_n x^n + u2_n x^{-(n+3)}\bigr] 
C_n^{3/2}(\mu)$,
where $x$ is non-dimensional radius,  $\mu = cos{\vartheta}$,
${\vartheta}$ is the co-latitude, $C_n^{3/2} {(\mu)}$ are the Gegenbaur
polynomials of order 3/2, $u1_n$ and $u2_n$
are the unknown constants. 
By taking a clue from the helioseismic inferences
 that meridional velocity increases from the surface towards base of the convective
envelope, we neglect first part
in the series solution and consider the second part $u2_n$ only.
Hence the required solution of the meridional
velocity in the convective envelope is given as
$U(x,\mu) = \bigl[(u2_1  x^{-4} C_1^{3/2}(\mu) + u2_3 x^{-6} C_3^{3/2}(\mu)\bigr]$. In order to solve
two unknown constants $u2_{1}$ and $u2_{3}$ uniquely, we match the observed surface meridional velocity
at the two latitudes $5^{o}$ and $60^{o}$ with the meridional
velocity obtained by the analytical solution.

The results show that meridional
velocity flow from the surface appears to penetrates deep below base of the 
convective envelope and at outer part of the radiative zone.
With such a deep flow velocity below the convective envelope and
a very high density stratification in the outer part of the
radiative zone with likely existence of a strong ($\sim$ $10^{4}$ G)
toroidal magnetic field structure, the velocity of transport of
meridional flow is considerably reduced.
Hence, it is very unlikely that the return flow will reach
the surface (with a period of solar cycle) as required
by some of the flux transport dynamo models.
On the other hand, deep meridional flow is required for
burning of Lithium at outer part of the radiative zone
supporting the observed Lithium deficiency at the surface.
\end{abstract}

\section{Introduction}

It is believed that meridional velocity flow may be transferring angular
momentum and maintaining solar differential rotation in the
convective envelope. In the early history of the stars, especially
the sun, meridional flow might have been played a major role in bringing
the lithium from the surface towards the outer part of the
radiative core where it is easily burnt and supplemented the existing
radiative energy flow. This may be one of the reason for the
 observed lithium depletion over the surface of the sun.

In case of the sun, the analysis from the tracers such as sunspots
 (Javaraiah 1999 and references there in; Wohl 2002),
magnetic field patterns (Snodgrass and Dailey 1996; Meunier 2005 and references there in),
 inferences from the Doppler measurements (Hathaway {\it et. al.}
1996; Nesme-Ribes {\it et. al.} 1997) and, the
inferences from the local helioseismology techniques (Giles {\it et. al.} 1997;
Chou and Dai 2001; Gizon, Duvall and Larsen 2001; Beck, Gizon and Duvall
2002; Basu and Antia 2002; Basu and Antia 2003;
Haber {\it et. al.} 2004; Gizon 2004; Zhao and Kosovichev 2004;
Antia and Basu 2007; Gizon and Thompson 2007;
Kriger, Roth and Luhe 2007; Shibahashi 2007)  
show that there exists a meridional velocity flow 
from the equator towards the pole. On the surface the flow 
velocity increases from $\sim$ 1-2 m/sec near the equator to
$\sim$ 20-50 m/sec near the higher latitudes. Some of the inferences
(Basu, Antia and Tripathy 1999; Antia and Basu 2007;
Gonzalez {\it et. al.} 2006) from the local helioseismology 
show that meridional flow increases from
surface towards base of the convective envelope. Unfortunately the helioseismic
inferences of meridional flow yield the accurate results only few mega
meters just below the surface. 

The genesis of meridional circulation in a star was first proposed by
Bierman (1958) who discussed extensively in the IAU
symposium that pure rotation without meridional circulation
is not possible stationary states of motion for the
convection zones of stars. Recent studies (Rudiger 1989;
Kitchatinov and Rudiger 2005; Tassoul 2000; Rempel 2005)
emphasize the turbulent Reynold stresses for maintaining
the differential rotation and hence existence of the
meridional flow. 

In order to reproduce proper solar butterfly diagrams and 
predict future solar cycles
(Dikpati and Gilman 2007; Choudhuri, Chatterjee and Jiang 2007),
 the flux transport dynamo models require the meridional
circulation that needs to penetrate (Nandy Choudhuri 2002)
 below base of the convective envelope. There are supporting  
(Rüdiger, Kitchatinov and Arlt 2005) and  nonsupporting
(Gilamn and Miesch 2004) models for the deep penetration
of the meridional velocity flow below base of the convective
envelope. Very recently Svanda, Zhao and Kosovichev (2007)
show that the mean longitudinally averaged meridional flow 
measurements by helioseismology may not be used directly in 
solar dynamo models for describing the magnetic flux transport, 
and that it is necessary to take into account the longitudinal 
structure of these flows. 

Aims of the present study are two fold: (i) with reasonable 
assumptions and approximations, solve modified Chandrasekhar's
MHD equation for meridional part of the velocity flow
in the convective envelope and, (ii) examine whether meridional 
velocity flow penetrate
deep below base of the convective envelope for the lithium burning
and also as required by the flux transport dynamo models.
Thus this study also supplements the information of the 
meridional flow velocity where local helioseismology can 
not infer reliably so deep below the surface.

\section {Reasonable assumptions and approximations}
As in our previous work (Hiremath \& Gokhale 1995; Hiremath 1994),
we assume that, in the convective envelope, the fluid is incompressible and the
large-scale magnetic fields and the fluid motions are symmetric about the rotation axis.
We also assume that the magnetic eddy diffusivity $\bf \eta$ and the eddy
diffusivity due to viscosity $\bf \nu$ are constants with values represented by
the appropriate averages.
                                                                                                   
        Following Chandrasekhar (1956), the magnetic field $\bf B$ and the
velocity $\bf V$ for the axisymmetric system can be expressed
\begin{equation}
{\bf h} = -{\varpi}{{\partial P}\over{\partial z}}{\bf \hat{\hbox{l}}}_\varpi + ({\varpi} T ){\bf \hat{\hbox{l}}}_\varphi+ {{1}\over{\varpi}}{{\partial}\over{\partial \varpi}}(\varpi^2 P)
%{\bf \hat{\hbox{l}}}_z  \, ,
{\bf \hat{\hbox{l}}}_z  \, ,
\end{equation}
\begin{equation}
{\bf V} = -{\varpi}{{\partial U}\over{\partial z}}{\bf \hat{\hbox{l}}}_\varpi
+({\varpi} \Omega){\bf \hat{\hbox{l}}}_\varphi
+ {{1}\over{\varpi}}{{\partial}\over{\partial \varpi}}(\varpi^2 U)
{\bf \hat{\hbox{l}}}_z \, ,
\end{equation}
\noindent where $ {\bf h} = {\bf B}/(4 \pi \rho)^{1/2}$,  $\rho$ is the density,
 ${\varpi}$, ${\varphi}$, $z$ are the cylindrical polar coordinates, with their
 axes along the axis of solar rotation; $\bf \hat{\hbox{l}}_\varpi$,
$\bf \hat{\hbox{l}}_\varphi $, and $\bf \hat{\hbox{l}}_z $ are the
corresponding unit vectors and; $P$, $T$, $\Omega$, and $U$ are the scalar
functions that are independent of $\varphi$.

Further we make the following assumptions and approximations.
                                                                                                   
  Steady parts of the poloidal magnetic field $P$ and poloidal
component of the velocity field $U$ (meridional velocity) are very weak
compared to the steady part of the rotation $\Omega$.
In fact such a steady part of poloidal magnetic field
is found to be $\sim 1$ G from the observation (Stenflo 1993) and $\sim 0.01$ G
from theoretical calculations (Hiremath and Gokhale 1995).
Thus , by taking average density
of the sun, Alfve$^{'}$n velocity varies between $\sim$ 1 - 0.01
cm $sec^{-1}$ which is very negligible compared to the
dominant part of rotational velocity ($\sim 10^{5}\, cm\,sec^{-1}$).
This leads us to safely assume
that $P$ is approximately zero and for the sake of making this investigation
simple we put $P=0$ in the following Chandrasekhar's MHD equations.
Similarly poloidal part of the velocity (meridional circulation)
over the surface is found to be $\sim\, 0.001 - 0.01 $ times
the rotation velocity. Though we can not neglect the meridional
velocity in MHD equations, it can not be equated with the 
 dominant part of the angular velocity $\Omega$.  We
also assume that strength of
steady part of toroidal field $T$ is less than (or at most comparable to)
that of the steady part of rotation. 

\section { Modified form of Chandrasekhar's equations}

The afore mentioned assumptions
lead to decoupling of poloidal part of velocity equation
and, thus we have the following modified Chandrasekhar's
(1956) MHD equations that take into account the eddy viscosity (Nakagawa and
Swarztrauber 1969) also

\begin{equation}
\nu\varpi\Delta_5({ \Delta_5 U}) + [\varpi^2 U, \Delta_5 U]
- \varpi{{\partial \Delta_5 U}\over{\partial t}} = 0 \, ,
\end{equation}
                                                                                                   
\begin{equation}
\eta\varpi\Delta_5 T + [\varpi^2 U, T]
- \varpi {{\partial T}\over{\partial t}} = 0 \, ,
\end{equation}
                                                                                                   
\begin{equation}
{\varpi^3}\nu \Delta_5 \Omega + [\varpi^2 U, \varpi^2 \Omega]
- {\varpi^3}{{\partial \Omega}\over{\partial t}} = 0 \, ,
\end{equation}

\begin{equation}
 {\varpi}{{\partial }\over{\partial z}}(T^2 - \Omega^2)  = 0 \, ,
\end{equation}
                                                                                                   
\noindent where
\begin{equation}
[f,g] = {{\partial f}\over{\partial z}}{{\partial g}\over{\partial \varpi}}
- {{\partial f}\over{\partial \varpi}}{{\partial g}\over{\partial z}} \,  ,
\end{equation} \nonumber
and
\begin{equation}
\Delta_5 = {{\partial^2}\over{\partial z^2}} + {3\over \varpi}{{\partial }
\over{\partial \varpi}} + {{\partial^2}\over{\partial \varpi^2}} \, . \nonumber
\end{equation}

In the previous study (Hiremath 2001), we used equations 4-6
to obtain the solution for the toroidal parts of magnetic
field and velocity field structures in the convective
envelope.  
\section{Solution and Results}
As the equation (3) for the meridional flow velocity $U$ 
is decoupled from rest of the 
equations, there are two unique solutions of this equation: either $U=0$ a
trivial solution and, $\Delta_5 U = 0$ a non trivial solution.
\begin{figure}[h]
\center{
  \noindent\includegraphics[width=25pc,height=20pc]{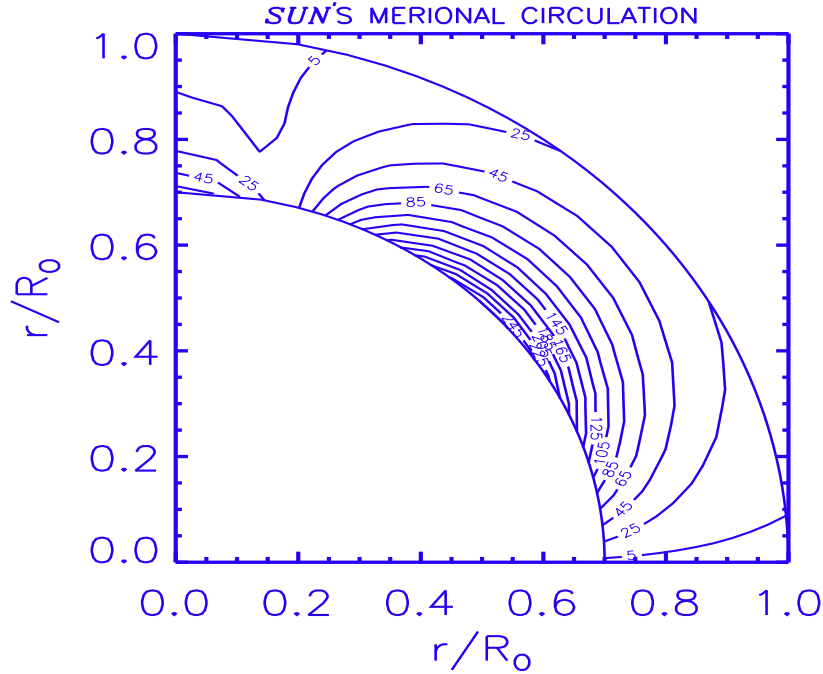}
}
\caption{Meridional circulation obtained from
the solution of Chandrasekhar's MHD equations in the sun's convective envelope}
\end{figure}

\noindent Nontrivial solution of the equation (3) is
$U(x,\mu) = \sum_{n=0}^\infty \bigl[u1_n x^n + u2_n x^{-(n+3)}\bigr] 
C_n^{3/2}(\mu)$,
where $x$ is non-dimensional radius,  $\mu = cos{\vartheta}$,
${\vartheta}$ is the co-latitude, $C_n^{3/2} {(\mu)}$ are the Gegenbaur
polynomials of order 3/2, $u1_n$ and $u2_n$
are the unknown constants to be determined from the boundary conditions.
By taking a clue from the helioseismic inferences 
(Basu, Antia and Tripathy 1999; Antia and Basu 2007;
Gonzalez {\it et. al.} 2006) that meridional
velocity increases from the surface towards base of the convective
envelope, we neglect first part
in the series solution and consider the second part $u2_n$ only.
For the sake of understanding and simplicity of the problem
we consider antisymmetric components $u2_{1}$ and $u2_{3}$ 
modes only. Hence the required solution of the meridional
velocity in the convective envelope is given as 
$U(x,\mu) = \bigl[(u2_1  x^{-4} C_1^{3/2}(\mu) + u2_3 x^{-6} C_3^{3/2}(\mu)\bigr]$. In order to solve
two unknown constants $u2_{1}$ and $u2_{3}$ uniquely, we match the observed surface meridional velocity
at the two latitudes $5^{o}$ and $60^{o}$ with the meridional 
velocity obtained by the analytical solution. 
Finally we get the meridional velocity $U$ from the determined
two unknown constants and using above equation. In Fig 1.,
we present the iso-meridional velocity flow in the
one quadrant of the convective envelope. The results show that: (i)
unlike the close isomeridional contours that are
required by the flux transport dynamo models, the present
solution yields the isomeridional contours that are not
closed in the convective envelope and appear to penetrate
deep in the outer part of the radiative zone, (ii) on the surface,
magnitude of meridional velocity is $\sim$ 5 m/sec
near the equator and reaches $\sim$ 25 m/sec around $45^{o}$
latitude and, (iii) near base of the convective envelope,
magnitude of meridional velocity is $\sim$ 5 m/sec
near the equator and reaches maximum of 
$\sim$ 220 m/sec around $45^{o}$ latitude. The first result
that meridional flow penetrates deeply in the outer part
of the radiative zone is also consistent with the result
obtained by the recent study (Garaud and Brummel 2007).  

Let us consider the isomeridional contour that is close to the surface
 ( $\sim$ 25 m/sec) and appears to penetrates deep in the outer
part of the radiative zone. Owing to very high 
density stratification in the outer part of the
radiative zone and likely existence (Friedland and Gruzinov 2004i;
 Rasba {\it et. al.} 2007)
of strong ($\sim$ $10^{4}$ G)
toroidal magnetic field structure, the velocity of transport of
meridional flow is considerably reduced (Rempel 2006).
The law of conservation of mass yields the relation
$U_{2}=(\rho_{1}/\rho_{2})U_{1}$, where $U_{1}$ and $\rho_{1}$
are meridional flow velocity and density stratification near the
surface and, $U_{2}$ and $\rho_{2}$ are meridional flow velocity and 
density stratification in the outer ($r/R_{\odot}$=0.5) part of 
the radiative zone. If we substitute $U_{1}=25$ m/sec and the
respective density values in this relation, meridional flow velocity
in the outer part of the radiative zone
is  found to be $\sim$ 1 cm/sec. If we assume that 
isomeridional velocity flows are circular loops, one
can notice from Fig 1 that half of the circular loop
is in the convective envelope and rest half lies in the
outer part of the radiative core. Thus travel time taken
by the longest meridional circular loop that lies in
the convective loop is found to be $\sim$ 3 years and the travel time taken
by the longest meridional circular loop that lies
in the outer part of the radiative region is found to be 
3000 years. That means the meridional   
flow velocity that starts near the surface and after penetrating
deep near base of the convective envelope return
back to the surface after 3000 years. This return
time scale is nearly 100 times the return time scale as
required by the flux transport dynamo models. {\it Hence it
is very unlikely that the solar meridional circulation
plays any major role in reproducing the proper
butterfly diagrams and dictating the next solar cycle}.

However, present study supports the idea that the observed 
surface Lithium depletion can be explained if deep meridional
circulation carries the Lithium and penetrates deep
in the outer part of the radiative core resulting in burning
 into helium.

%\ack
%\medskip

\smallskip

\end{document}